\documentstyle[12pt]{article}
\begin{document}
\renewcommand{\thetable}{\Roman{table}}
\newcommand{\be}{\begin{eqnarray}}
\newcommand{\beq}{\begin{equation}}
\newcommand{\ba}{\begin{array}}
\newcommand{\ee}{\end{eqnarray}}
\newcommand{\eeq}{\end{equation}}
\newcommand{\ea}{\end{array}}
\newcommand{\zt}{\zeta}
\newcommand{\ve}{\varepsilon}
\newcommand{\al}{\alpha}
\newcommand{\gm}{\gamma}
\newcommand{\Gm}{\Gamma}
\newcommand{\om}{\omega}
\newcommand{\et}{\eta}
\newcommand{\bt}{\beta}
\newcommand{\dt}{\delta}
\newcommand{\Dt}{\Delta}
\newcommand{\La}{\Lambda}
\newcommand{\la}{\lambda}
\newcommand{\vp}{\varphi}
\newcommand{\nn}{\nonumber}
\newcommand{\nid}{\noindent}
\newcommand{\lmx}[1]{\begin{displaymath} {#1}=
                    \left(\begin{array}{rrr}}
\newcommand{\rmx}{\end{array} \right) \end{displaymath}}

\begin{titlepage}
\begin{center}
 {\Large \bf
 Critical thermodynamics of three-dimensional $MN$-component field
 model with cubic anisotropy from higher-loop $\ve$ expansion}

\vspace{1cm}
 {\Large A. I. Mudrov$^*$,  K. B. Varnashev$^{**}$}

\bigskip
{\it $^{*}$ Department of Mathematics, Bar-Ilan University,
52900 Ramat-Gan, Israel; \\
e-mail: mudrova@macs.biu.ac.il \\
$^{**}$
Department of Physical Electronics, Saint Petersburg
Electrotechnical University, \\
Professor Popov Street  5, St. Petersburg, 197376, Russia; \\
e-mail: kvarnash@kv8100.spb.edu}
\end{center}
\vspace{0.25cm}
\begin{abstract}
\vspace{0.75cm}
The critical thermodynamics of an $MN$-component field model with
cubic anisotropy relevant to the phase transitions in certain crystals 
with complicated ordering is studied within the four-loop $\ve$ expansion
using the minimal subtraction scheme. Investigation of the global structure 
of RG flows for the physically significant cases $M=2$, $N=2$ and $M=2$, $N=3$ 
shows that the model has an anisotropic stable fixed point with new critical 
exponents. The critical dimensionality of the order parameter is proved to be 
equal to $N_c^C=1.445(20)$, that is exactly half its counterpart in the real 
hypercubic model.
\end{abstract}

\vspace{0.5cm}
\vspace{1cm}

\nid
~\qquad Published in {\sl J. Phys. A: Math. \& Gen.} {\bf 34} (2001) L347-L353.

\vspace{7cm}

\rightline{{\sl Typeset using} \LaTeX}
\end{titlepage}

\newpage
We study the critical behaviour of an $MN$-component
field model with cubic anisotropy having a number of interesting
applications to phase transitions in three-dimensional simple and
complicated systems. The effective Ginzburg-Landau Hamiltonian 
of the model reads:
\be
H =
\int d^{~d}x \Bigl[{1 \over 2} \sum_{\al=1}^N (m_0^2~ |\vec \vp^{\al}|^2
 + |\nabla \vec \vp^{\al}|^2)
 + {u_0 \over 4!} \Bigl(\sum_{\al=1}^N |\vec \vp^{\al}|^2 \Bigr)^2
 +  {v_0 \over 4!} \sum_{\al=1}^N |\vec \vp^{\al}|^4 \Bigr]
\label{eq:Ham}
\ee
where each vector field $\vec \vp^\al$ has $M$ real 
components\footnote{For $M=1$ and $M=2$ the model (\ref{eq:Ham}) is 
merely the $N$-component cubic model determined either by the real or by 
the complex order parameter field, respectively.} $\vp^\al_i$,
$i=1, \ldots, M$ and $d=4 - \ve$ is the spatial dimensionality. 
Here $m_0^2 \sim (T-T_c)$ and $m_0$, $u_0$, $v_0$ are the "bare" mass 
and coupling constants, respectively.

For $M=N=2$ the  Hamiltonian (\ref{eq:Ham}) describes the structural phase
transition in NbO$_2$ crystal and the antiferromagnetic phase transitions
in TbAu$_2$ and DyC$_2$. Another physically important case $M=2$, $N=3$ is
relevant to the antiferromagnetic phase transitions in K$_2$IrCl$_6$, 
TbD$_2$ and Nd materials \cite{Muk}. The magnetic and
structural phase transitions in a cubic crystal are governed by the model
(\ref{eq:Ham}) at $M=1$ and $N=3$ \cite{Ah73}.
In the replica limit $N \to 0$ ($M=1$) the
Hamiltonian (\ref{eq:Ham}) is known to determine the critical properties of
weakly disordered quenched systems undergoing second-order phase
transitions \cite{GrLutAh} with a specific set of critical exponents
\cite{HLKh}. Finally, the case $M=1$ and $N \to \infty$ corresponds to the
Ising model with equilibrium magnetic impurities \cite{Ah73l}. In this limit
the Ising critical exponents take Fisher-renormalized values \cite{F68}.
Since  static critical phenomena in a cubic crystal as well as in randomly
diluted Ising spin systems are well understood
\cite{KSf,KTSf,VKB,CPV,FHY}\footnote{For a six-loop study of the critical 
behaviour of the random Ising model see \cite{Vic}}, we will focus here on 
the critical behaviour
of the above mentioned multisublattice antiferromagnets. This is the case of
$M=2$ and $N=2, N=3$ in the fluctuation Hamiltonian (\ref{eq:Ham}).

For the first time,  magnetic and structural phase transitions in crystals
with complicated ordering described by the model (\ref{eq:Ham}) were
studied by Mukamel and Krinsky within the lowest orders in $\ve$ \cite{Muk}.
A stable fixed point ("unique" point), different from the isotropic 
($O(MN)$-symmetric) or the Bose ($O(MN)$-symmetric, $MN=2$) one\footnote{Note 
that the isotropic or the Heisenberg fixed point can be determined as the 
point having the coordinates $u>0$, $v=0$ on the RG flow diagram, whereas 
for the Bose fixed point one has $u=0$, $v>0$.}, was predicted in $d=3$. 
The point was 
shown to determine a new universality class with a specific set of critical 
exponents. However, for the physically important case $M=N=2$, the critical 
exponents of the unique fixed point turned out to be the same as those of the 
$O(4)$-symmetric one within the two-loop approximation.

Later an alternative analysis of critical behaviour of the model, the
RG approach in fixed dimension, was carried out within the two- and
three-loop approximations \cite{Shp,VS}. Those investigations gave the same
qualitative predictions: the unique stable fixed point does exist on the 
three-dimensional RG flow diagram. 
However, the critical exponents computed at this point with
the use of different resummation procedures proved to be close to those
of the Bose fixed point rather than the isotropic one. It was also shown
that the unique and the Bose fixed points are very close to each other on the
diagram of RG flows, so that they may interchange their stability in the next
orders of RG approximations \cite{VS}.

Recently, the  critical properties of the model were
analyzed in third order in $\ve$ \cite{BGMV,MV-1}. Investigation of
the fixed point stability and calculation of the critical dimensionality $N_c$
of the order parameter, separating two different regimes of critical behaviour,
confirmed that the model (\ref{eq:Ham}) possesses the anisotropic (complex 
cubic; $M=2$ and $u\ne0$, $v\ne0$) stable fixed point for $N=2$ and 3. 
The realistic critical exponent estimates for the unique stable fixed point 
were obtained in \cite{MV-1} using the summation technique of \cite{MV-2}.
The values appeared to be close to those of the isotropic point in 
contradiction to the numerical results given by RG directly in three 
dimensions \cite{Shp,VS}. Such a distinction may be accounted for by the too 
short three-loop $\ve$ series used.

It is worthy of note that the existence of an anisotropic stable fixed point
on the three-dimensional RG flow diagram contradicts the nonperturbative 
considerations \cite{CB}. 
Indeed, according to those considerations the only stable fixed point in three 
dimensions may be the Bose one and it is that point which governs the 
critical thermodynamics in the phase transitions of interest. The point is 
that the model (\ref{eq:Ham}) describes $N$ interacting Bose systems.
As was shown by Sak \cite{Sak}, the interaction term can be
represented as the product of the energy operators of various two-component
subsystems. It was also found that one (the smallest) of the eigenvalue exponents 
characterizing the evolution of this term under the renormalization group
in a neighborhood of the Bose fixed point is proportional to the specific heat
exponent $\al$. Since $\al$ is believed to be negative at that point, and
that is confirmed experimentally \cite{Lipa} and theoretically
\cite{GdZ98}, the interaction is irrelevant. Consequently, the Bose fixed
point should be stable in three dimensions. However, the RG approach, applied 
to the model (\ref{eq:Ham}), has not yet confirmed this conclusion.
It is therefore highly desirable to extend already known $\ve$ expansions
for the stability matrix eigenvalues, critical exponents and the critical 
dimensionality in order to apply  more sophisticated resummation techniques 
to longer expansions.

Therefore, the aim of this Letter is to extend the existing three-loop 
$\ve$ expansions of the model to the four-loop order and to study more 
carefully the predictions of the RG method regarding the critical phenomena 
in the substances of interest.
Namely, on the basis of the four-loop expansions for the RG functions obtained
using dimensional regularization and the minimal subtraction scheme
\cite{Hooft,Br}, we analyze the stability of the fixed points and calculate
the critical dimensionality of the order parameter. We show that the 
anisotropic stable fixed point does exist on the thre-dimensional RG flow 
diagram. For this point, we give more accurate critical exponent estimates, in 
comparison with the previous three-loop results \cite{MV-1}, by applying a 
new summation approach \cite{MV-2} to the four-loop series.

The four-loop $\ve$ expansions for the $\bt$-functions of the model are
as follows
\be
\bt_u &=& \ve u -u^2 - \frac{4}{N + 4} u v
 + \frac{1}{(N+4)^2} \Bigl[3 u^3 (3 N + 7) + 44 u^2 v + 10 u v^2 \Bigr]
\label{eq:Bu} \\
&-& \frac{1}{(N + 4)^3} \Bigl[\frac{u^4}{4}(48 \zt(3) (5 N+11)
 +33 N^2+461 N+740) + u^3 v (384 \zt(3) +79 N
\nn \\
&+& 659) + \frac{ u^2 v^2}{2} (288 \zt(3) +3 N +1078) + 141 u v^3 \Bigr]
 - \frac{1}{(N + 4)^4} \Bigl[\frac{u^5}{12}(- 48 \zt(3) (63 N^2
\nn \\
&+& 382 N + 583) + 144  \zt(4)( 5 N^2+31 N +44) - 480 \zt(5) (4 N^2
 + 55 N + 93)+ 5N^3
\nn \\
&-& 3160 N^2 - 20114 N- 24581) - \frac{2 u^4 v}{3} (12 \zt(3) (3 N^2
 + 276 N + 1214) - 36 \zt(4)
\nn \\
&\times& (19 N + 85) + \zt(5) (2400 N + 23040) - 28 N^2 + 3957 N +15967)
 - \frac{u^3 v^2}{3} (72 \zt(3)
\nn \\
&\times& (19 N + 426) - 4032 \zt(4) + 39840 \zt(5) + 1302 N + 46447)
 + \frac{2 u^2 v^3}{3} (60 \zt(3) (N
\nn \\
&-& 84) - 792 \zt(4) - 4800 \zt(5) - 125 N - 12809)
 - \frac{u v^4}{2} (400 \zt(3) + 768 \zt(4) + 3851) \Bigr]
\nn \\
\bt_v &=& \ve v - \frac{1}{N+4}(6 u v + 5 v^2) + \frac{1}{(N+4)^2}
\Bigl[u^2 v ( 5 N + 41) + 80 u v^2 + 30 v^3 \Bigr]
\label{eq:Bv} \\
&-& \frac{1}{(N + 4)^3} \Bigl[\frac{u^3 v}{2} (96 \zt(3) (N + 7)
 - 13 N^2 + 184 N + 821) + \frac{u^2 v^2}{4} (4032 \zt(3) + 59 N
\nn \\
&+& 5183) + u v^3 (768 \zt(3) + 1093) + \frac{v^4}{2} (384 \zt(3) + 617) \Bigr]
 - \frac{1}{(N + 4)^4} \Bigl[\frac{u^4 v}{4} (48 \zt(3)
\nn \\
&\times& (N^3 -12 N^2 - 140 N - 567) + 144 \zt(4) (2 N^2 + 17 N +45)
-3360 \zt(5) (3 N + 13)
\nn \\
&-& 29 N^3 - 28 N^2 - 6958 N - 19679)
 + \frac{u^3 v^2}{3} (12 \zt(3) (9 N^2 - 591 N -7028) + \zt(4)
\nn \\
&\times& (3528 N + 21240) - 480 \zt(5)(10 N + 287) + 61 N^2 - 5173 N - 66764)
- \frac{u^2 v^3}{3}
\nn \\
&\times& (1800 \zt(3) (N + 62) - 144 \zt(4) (8 N + 203)
 + 172800 \zt(5) + 56 N + 93701)
\nn \\
&-& 4 u v^4 (5090 \zt(3) - 1296 \zt(4) + 7600 \zt(5) + 4503)
 + \frac{v^5}{2} (- 8224 \zt(3) + 1920\zt(4)
\nn \\
&-& 12160 \zt(5)-7975) \Bigr] \nn
\ee
where $\zt(3)$,$\zt(4)$, and $\zt(5)$ are the Riemann $\zt$ functions.

From the system of equations $\bt_u(u^*, v^*)=0$ and $\bt_v(u^*, v^*)=0$,
we found formal series for the four fixed points:  trivial Gaussian 
and nontrivial isotropic, Bose, and complex cubic.
Then we calculated the eigenvalues of the stability matrix
\lmx{\Omega}
    \frac{\partial\bt_u(u,v)}{\partial u} &
    \frac{\partial\bt_u(u,v)}{\partial v} \\
    \frac{\partial\bt_v(u,v)}{\partial u} &
    \frac{\partial\bt_v(u,v)}{\partial v}
\rmx
taken at the most intriguing Bose and complex cubic fixed points.
The corresponding numerical estimates are obtained using an approach
based on the Borel transformation 
\be
 F(\ve;a,b) =\sum_{k=0}^\infty A_k(\la) \int_0^\infty
 e^{-\frac{x}{a \ve}} \Bigl(\frac{x}{a \ve}\Bigr)^b
 d\Bigl(\frac{x}{a \ve}\Bigr)
 \frac{z^k(x)}{[1-z(x)]^{2\la}}
 \label{eq:Bor}
\ee
modified with a conformal mapping $z=\frac{\sqrt{x+1}-1}{\sqrt{x+1}+1}$
\cite{LgZ}, which does not require the knowledge of the exact asymptotic
high-order behaviour of the series \cite{MV-2}. The parameter $\la$ is 
chosen from the condition of the most rapid convergence of series 
(\ref{eq:Bor}), that is by minimizing the quantity 
$|1-\frac{F_l(\ve;a,b)}{F_{l-1}(\ve;a,b)}|$, where $l$ is the step of 
truncation and $F_l(\ve;a,b)$ is the $l$-partial sum for $F(\ve;a,b)$. 
If the real parts of
both eigenvalues are negative, the associated fixed point is infrared stable 
and the critical behaviour of experimental systems undergoing second-order 
transitions is determined only by that stable point. For the Bose and
the complex cubic fixed points our numerical results are presented in table 1.
It is seen that the complex cubic fixed point is absolutely stable in $d=3$
($\ve=1$), while the Bose point appears to be of the "saddle"
type\footnote{We say the fixed point is of the "saddle" type provided 
their eigenvalue exponents $\om_1$ and $\om_2$ are of 
opposite sings in the ($u, v$) plane.}.
However, the $\om_2$ of either point is very small at the four-loop level,
thus implying that these points may swap their stability in the next order
of the RG approximation.

In addition to the eigenvalues, we calculated the critical dimensionality
$N_c^C$ of the order parameter from the condition of vanishing $\om_2$
for the complex cubic fixed point. 
The four-loop expansion is
\be
N_c^C = 2 - \ve + \frac{5}{24} \Bigl[6 \zt(3) -1 \Bigr] \ve^2
 + \frac{1}{144} \Bigl[45 \zt(3) + 135 \zt(4) - 600 \zt(5) -1 \Bigr] \ve^3.
\label{eq:Nc}
\ee
Instead of processing this expression numerically,
we established the exact relation $N_c^C=\frac{1}{2} N_c^R$, which is
independent on the order of approximation used\footnote{Here $N_c^C$ and
$N_c^R$ are the critical (marginal) spin dimensionalities in the complex 
($M=2$) and in the real ($M=1$) hypercubic model, respectively.}.
In fact, the critical dimensionality $N_c^C$ for the complex cubic model is
determined as that value of $N^C$ at  which the complex cubic fixed point
coincides with the isotropic one. The same assertion holds for the cubic model
with the real $N^R$-component order parameter. So, because of the relation
$O(2N^C)=O(N^R)$, the relation $2 N_c^C=N_c^R$ should hold too.

The five-loop $\ve$-expansion for $N_c^R$ was recently obtained in Ref.
\cite{KSf}.
Resummation of that series gave the estimate $N_c^R=2.894(40)$ \cite{VKB}.
Therefore we conclude that $N_c^C=1.447(20)$ from the five-loops.
Practically the same estimate $N_c^C=1.435(25)$ follows from a
constrained analysis of $N_c^R$ taking into account $N_c^R=2$ 
in two dimensions \cite{CPV}. From the recent pseudo-$\ve$ expansion
analysis of the real hypercubic model \cite{FHY} one can extract
$N_c^C=1.431(3)$. However the most accurate estimate
$N_c^C=1.445(20)$ results from the value $N_c^R=2.89(4)$ obtained on the
basis of the numerical analysis of the four-loop \cite{VKB} and the six-loop
\cite{CPV} three-dimensional RG expansions for the $\bt$-functions of the real 
hypercubic model. Since $N_c^C < 2$, the phase transitions
in the NbO$_2$ crystal and in the antiferromagnets TbAu$_2$,
DyC$_2$, K$_2$IrCl$_6$, TbD$_2$ and Nd are of second order and their
critical thermodynamics should be controlled by the complex cubic fixed point
with a specific set of critical exponents, in the frame of the given 
approximation.
The corresponding four-loop critical exponent estimates are displayed in 
table 2. The critical exponent estimates obtained for the isotropic and 
the Bose fixed points are also presented in the table, for comparison.

In conclusion, the four-loop $\ve$-expansion analysis of the generalized
$MN$-component Ginzburg-Landau model with cubic anisotropy describing phase
transitions in certain real antiferromagnets with complicated ordering has
been carried out with the use of the minimal subtraction scheme.
Investigation of the global structure of RG flows for the physically
significant cases $M=2$, $N=2$ and $M=2$, $N=3$ has shown that the complex
cubic rather than the Bose fixed point is absolutely stable in three 
dimensions.
The critical dimensionality $N_c^C=1.445(20)$ of the order parameter obtained
from  six loops has confirmed this conclusion. For the stable complex
cubic fixed point, reasonable estimates of critical exponents were obtained
using the Borel summation technique in combination with a conformal
mapping. For the structural phase transition in NbO$_2$ and for the
antiferromagnetic phase transitions in TbAu$_2$ and DyC$_2$, they were shown
to be close to the critical exponents of the O(4)-symmetric model. 
In contrast to this, the critical exponents
for the antiferromagnetic phase transitions in K$_2$IrCl$_6$, TbD$_2$ and Nd
turned out to be close to the Bose ones. Although our results seem to be 
substantially consistent with other predictions of the RG approach, still 
there is a definite contradiction with the general nonperturbative theoretical 
arguments \cite{CB} mentioned in the introduction. 
One can hope, however, that the
five-loop contributions being taken into account will eliminate this
controversy. Indeed, the present calculations have shown that,
although the complex cubic fixed point, rather than the Bose one, is stable
at the four-loop level, the eigenvalues $\om_2$ of both fixed 
points are very small.
Therefore the situation is very close to marginal, and
the points might change their stability to the opposite in the next order of
perturbation theory, so the Bose point would turn out to be stable. 
There is a hope that comparison of the critical exponent values 
obtained theoretically for different fixed points with those values 
determined from experiments or, probably, from Monte Carlo simulations
would indicate which fixed point is really stable in three dimensions.
Finally, it would also be desirable to investigate certain universal amplitude 
ratios of the model because they vary much more among different 
universality classes than exponents do and might be more effective as a 
diagnostic tool.

We are grateful to Professor M. Henkel for helpful remarks and to 
Dr. E. Blagoeva for communicating her results mentioned in \cite{BGMV}. 
One of the authors (K.B.V.) acknowledges useful discussions with 
Dr. B. N. Shalaev. 
This work was supported by the Russian Foundation for Basic Research, grant
no 01-02-17048, and by the Ministry of Education of Russian Federation, grant 
no E00-3.2-132.

\bigskip
\nid
{\it Note added.}
More recently, the six-loop RG functions of the model of interest have been
calculated within the alternative RG approach in three dimensions 
\cite{Vic,PV-z}. Although 
the authors argue the global stability of the Bose fixed point, the numerical
estimate of the smallest stability matrix eigenvalue of the Bose
fixed point, which governs the RG flows near this point,
appears to be very small, $\om_2=-0.007(8)$ \cite{PV-z}, and
the apparent accuracy of the analysis does not exclude the opposite sing for
$\om_2$. This result agrees well with our conclusions. 

\begin{table}
\caption{Eigenvalue exponents estimates obtained for the Bose (BFP) 
and the complex cubic (CCFP) fixed points at $N=2$ and $N=3$ within the
four-loop approximation in $\ve$ ($\ve=1$) using Borel transformation with
a conformal mapping.}
\label{TabI}
\vspace{0.5cm}
\hspace{1.5cm}
\begin{tabular}{|c|l|l|l|l|}\hline
Type of             &\multicolumn{2}{|c|}{$N=2$}&
                     \multicolumn{2}{|c|}{$N=3$}
                   \\[0pt]    \cline{2-5}
fixed point         &\multicolumn{1}{|c|}{$\om_1$}
                    &\multicolumn{1}{|c|}{$\om_2$}
                    &\multicolumn{1}{|c|}{$\om_1$}
                    &\multicolumn{1}{|c|}{$\om_2$}
                   \\[0pt] \hline
BFP                 &$-0.395(25)$ & $0.004(5)$
                    &$-0.395(25)$ & $0.004(5)$
                   \\[0pt] \hline
CCFP                &$-0.392(30)$ & $-0.029(20)$
                    &$-0.400(30)$ & $-0.015(6)$
                   \\[0pt]\hline
\end{tabular}
\end{table}

\begin{table}
\caption{Critical exponents calculated for the isotropic (IFP), the Bose (BFP),
and the complex cubic (CCFP) fixed points at $N=2$ and $N=3$ within the
four-loop approximation in $\ve$ ($\ve=1$) using Borel transformation with a conformal
mapping.}
\label{TabII}
\vspace{0.5cm}
\begin{tabular}{|c|l|l|l|l|l|l|}\hline
Type of             &\multicolumn{3}{|c|}{$N=2$}&
                     \multicolumn{3}{|c|}{$N=3$}
                   \\[0pt]    \cline{2-7}
fixed point         &\multicolumn{1}{|c|}{$\eta$}
                    &\multicolumn{1}{|c|}{$\nu$}
                    &\multicolumn{1}{|c|}{$\gm$}
                    &\multicolumn{1}{|c|}{$\eta$}
                    &\multicolumn{1}{|c|}{$\nu$}
                    &\multicolumn{1}{|c|}{$\gm$}
                   \\[0pt]  \hline
IFP                 &$0.0343(20)$ & $0.725(15)$ & $1.429(20)$
                    &$0.0317(10)$ & $0.775(15)$ & $1.524(25)$
                   \\[0pt] \hline
BFP                 &$0.0348(10)$ & $0.664(7)$ & $1.309(10)$
                    &$0.0348(10)$ & $0.664(7)$ & $1.309(10)$
                   \\[0pt] \hline
CCFP                &$0.0343(20)$ & $0.715(10)$ & $1.404(25)$
                    &$0.0345(15)$ & $0.702(10)$ & $1.390(25)$
                   \\[0pt]\hline
\end{tabular}
\end{table}

\nid

\end{document}